\documentclass{aa} 
\usepackage{psfig}
\def\be {\begin{equation}}
\def\ee {\end{equation}}
\def\simless{\mathbin{\lower 3pt\hbox
{$\rlap{\raise 5pt\hbox{$\char'074$}}\mathchar"7218$}}} 
\def\simgreat{\mathbin{\lower 3pt\hbox
{$\rlap{\raise 5pt\hbox{$\char'076$}}\mathchar"7218$}}} 

\def\adhoc {{\it ad hoc}}

\def\v2  {v$_2$}

\def\mic {$\mu\hbox{m}$}

\def\kms {\hbox{${\rm km\ts s}^{-1}$}}
\def\percc {$\hbox{{\rm cm}}^{-3}$}    
\def\cmsq  {$\hbox{{\rm cm}}^{-2}$}    
\def\arcsec {\hbox{$^{\prime\prime}$}}

\begin{document}

%
\title{The dust temperature distribution in prestellar cores}


\author{A.~Zucconi \inst{1}
\and
C.M.~Walmsley \inst{2}
\and
D.~Galli \inst{2}
}

\offprints{M.~Walmsley}
\mail{walmsley@arcetri.astro.it}

\institute{Dipartimento di Astronomia e Scienza dello Spazio, Universit\`a di Firenze,
Largo E.~Fermi 5, I-50125 Firenze, Italy \\
email: zucconi@arcetri.astro.it
\and
Osservatorio Astrofisico di Arcetri, Largo E.~Fermi 5,
I-50125 Firenze, Italy\\
email: walmsley@arcetri.astro.it, galli@arcetri.astro.it
}

\date{Received ; accepted }

\abstract{We
 have computed the dust temperature distribution to be expected in a
pre--protostellar core in the phase prior to the onset of gravitational
instability. We have done this in the approximation that the heating of
the dust grains is solely due to the attenuated external radiation field
and that the core is optically thin to its own radiation. This permits us
to consider non spherically symmetric geometries. We predict the intensity
distributions of our model cores at millimeter and sub--millimeter
wavelengths and compare with observations of the well studied object L1544.
We have also developed an analytical approximation for the temperature at
the center of spherically symmetric cores and we compare this with the
numerical calculations.  Our results show (in agreement with Evans et al.
2001) that the temperatures in the nuclei of cores of high visual extinction
($>$ 30 magnitudes) are reduced to values of below $\sim 8$~K or roughly half
of the surface temperature. This has the consequence that maps at wavelengths
shortward of 1.3 mm see predominantly the low density exterior of
pre--protostellar cores. It is extremely difficult to deduce the true
density distribution from such maps alone. We have computed the intensity
distribution expected on the basis of the models of Ciolek \& Basu~(2000)
and compared with the observations of L1544.  The agreement is good
with a preference for higher inclinations (37$^{\circ}$ instead of
$16^{\circ}$ ) than that adopted by Ciolek \& Basu~(2000). We find that
a simple extension of the analytic approximation  allows a reasonably
accurate calculation of the dust temperature as a function of radius
in cores with density distributions approximating those expected for
Bonnor--Ebert spheres and suggest that this may be a useful tool for
future calculations of the gas temperature in such cores.
}
\maketitle

\section{Introduction}

The structure of the high density cores embedded in nearby molecular
clouds is currently a matter of great interest. The fate of a protostar
depends sensitively on the initial conditions prior to  collapse and
it is difficult to infer this from studies of ``protostars'' where a
luminous object has already formed. Considerable observational effort
has therefore been expended in the attempt to detect a
``pre-protostellar core'', which is taken to be a cold dense core whose
column density and internal pressure exceed considerably those found in
the surroundings. The high density nucleus of the dark cloud L1544 is
an example of such an object (Tafalla et al.  1998, Ward--Thompson et
al. 1999, Williams et al. 1999, Ciolek \& Basu 2000a,b) and in fact the
models of Ciolek \& Basu (2000a, CB hereafter) suggest that it will
become unstable in around 30000 years from now.  However, these inferences
depend on our ability to derive the density distribution  in such
sources from observables such as  the millimeter emission and the
mid--IR absorption (e.g. Bacmann et al. 2000).

In particular, the emission of dust grains at millimeter and
submillimeter wavelengths is sensitively dependent not only on the
grain column density but also upon the grain temperature.  In clouds
such as L1544, the grain temperature is known  to be of order 12~K on
the basis of the observed spectral energy distribution (Andr\'{e} et
al.~2000) and it has been normal practise to assume such regions to be
isothermal. However, it has been known for some time that one can
expect appreciable temperature gradients within cores heated by the
(external) interstellar radiation field (e.g. Leung 1975;
Mathis et al.~1983, hereafter MMP) as
well as temperature differences between grains of differing optical
properties.  For example, for silicate grains, MMP
predict grain temperatures as low as 6~K in the nucleus of cores of
visual extinction $\sim$ 50 magnitudes.  Such low temperature grains will
contribute negligibly to the emission at wavelengths below 1~mm and
this suggests that a new study of grain temperatures in dense cloud
cores similar to L1544 is warranted.

Another characteristic of the observed pre--protostellar cores is that
they clearly show large departures from spherical symmetry in both the
maps of dust emission and absorption (Bacmann et al. 2000).  This is
often interpreted as being due to flattening along the direction of the
mean magnetic field resulting in a disk-like configuration (Ciolek \&
Basu 2000a,b). Furthermore, as shown recently by Galli et al.~(2001), a
cloud core modeled as a thin disk perpendicular to the large-scale
magnetic field and supported against its self-gravity by a combination
of gas pressure and magnetic forces, is not necessarily axisymmetric.
Direct evidence for the importance of a magnetic field in determining
the structure of L1544 comes both from polarization measurements at
850 \mic \ (Ward--Thompson et al. 2000) and observation of OH Zeeman
splitting (Crutcher \& Troland 2000). The magnetic field inferred from
the Zeeman measurement is consistent with the model predictions of CB
but this result is sensitive to the assumed inclination of the magnetic
field relative to the plane of the sky (16$^{\circ}$, according to CB).
It is
relevant also that the  observed direction of polarization is {\it not}
consistent with naive CB model expectations in that the inferred
magnetic field direction in the plane of the sky deviates by an angle
of 30$^{\circ}$ from that expected (parallel to the L1544 minor
axis).  This can be easily explained if the cloud is slightly
non-axisymmetric (Basu~2000, Galli et al. 2001). Thus, while it seems
plausible that the magnetic field plays an important role in the
evolution of pre--protostellar cores just prior to collapse, it is
possible that the ``standard'' ambipolar diffusion model requires
modification.  One notes however that the effects of temperature
gradients mentioned earlier will have the qualitative effect of biasing
the 850 \mic \ polarization measurements towards the lower density
outer parts of cores such as L1544.  Thus it is clearly of importance
to be able to assess how large such temperature gradients really are.

With this  in mind, we report here new calculations of the expected
temperature distribution in pre--protostellar cores. We note  that a
rather similar study has recently been carried out by Evans et al.
(2001, hereafter ERSM).  These authors have confined themselves to
models with spherical symmetry  and have compared their model
predictions to the observations  of Shirley et al.~(2000). We, in
contrast, have employed a technique which allows us to simulate non
spherically symmetric situations such as the magnetic field dominated
models mentioned earlier.  We however concur with ERSM in their
conclusion that the decrease of temperature towards the centers of
cores strongly affects the observed sub--millimeter emission. We also
use our models to predict the emission distribution expected on the
basis of the CB models and conclude that a reasonable fit to the
observations can be obtained.

The structure of this paper is as follows. In section 2, we describe
the model which we have used and the assumptions which we have made. In
section 3, we present an analytic treatment aimed at determining
the dust temperature at the center of a spherically symmetric cloud.
In section 4, we give numerical results for both spherically symmetric models
and models based on the density distributions of the CB models.
Then, in
section 5, we  consider the intensity distributions which would
be expected for mm--submm maps of our model cores and
compare with observed data.  In section 6, we summarize our
conclusions.

\section{Model}

Our model calculations are aimed at calculating grain temperatures
under physical conditions similar to those  thought to pertain in
L1544 and similar pre--protostellar cores. We have made a number of
simplifying assumptions in doing this which we justify in the following
discussion. One of these is that we neglect scattering as we are interested
in studying cores of more than 10 magnitudes of visual extinction which are
mainly penetrated by the external infrared interstellar radiation field
for which the Rayleigh limit holds and thus scattering becomes negligible.

Another important assumption which we make 
is that  grains are only heated by the incident radiation
field and that other sources of grain heating are negligible.  ERSM
provide a detailed discussion of the range of applicability of this and
conclude that for representative conditions with  dust grain
temperatures above 5~K, heating due to the ambient radiation field
dominates.  We follow them in this but differ from them in {\it only}
considering heating due to the (attenuated) external radiation field.
That is to say, we assume that our model cores are optically thin to
their own radiation and that we can neglect re--absorption of radiation
emitted from within our model itself.  This is an essential
simplification  which permits us to consider non
spherically symmetric geometries.  However, it  restricts us to
considering structures with hydrogen column densities  less than a few
hundred  visual magnitudes of extinction.

The rationale behind this can be understood if one notes that for
typical temperatures of order 10~K, one expects the bulk of the
emission from a pre--protostellar core such as L1544 at wavelengths of
order 200--300~\mic.  The available observations (see Andr\'{e} et al.~2000
for the case of L1544) fully confirm this. Then to maintain the
optical depth below, say, 0.1 at 200~\mic \ requires a column density
of ${\rm H_2}$ less than $5\times10^{23}$ \cmsq \ corresponding to
roughly 500
magnitudes of visual extinction.  Available observational data suggest
that this condition is satisfied in all cases.  We note however that
for models with a singularity at the origin such as the singular
isothermal sphere of Shu~(1977), this condition is strictly
speaking not fulfilled.

We determine the grain temperature at a given position  simply by
applying the ``classical'' equilibrium between grain cooling and
heating for a spherical grain at position $r$ in the dust
cloud (see e.g. Spitzer 1978 or Boulanger et al. 1998 for a
discussion),
\be
\int^{\infty}_{0} Q_\nu B_\nu[T_{\rm d}(r)]\; d\nu =
\int^{\infty}_{0} Q_\nu J_{\nu}(r) \; d\nu.
\label{therm_eq}
\ee
Here, the right hand side describes the grain heating due to an incident
radiation field of average intensity $J_{\nu}(r)$ upon grains of
absorption efficiency $Q_\nu$ at frequency $\nu$.  The left hand side
gives the cooling rate for a grain of temperature $T_{\rm d}(r)$ and
$B_\nu$ is the Planck function.

The only real computational difficulty here is posed by the incident
radiation field $J_{\nu}(r)$ which we suppose  to be given by the
attenuated interstellar radiation field $J_\nu^{\rm is}$.  Thus, we have :
\be
J_{\nu}(r) =  \frac{J_\nu^{\rm is}}{4\pi}\int
\exp [-\tau_\nu(r,\theta,\phi)]\; d\Omega
\label{Jrdef}
\ee
Here, the integral over solid angle describes the average attenuation
due to an optical depth $\tau_\nu(r,\theta,\phi)$ in any given
direction.

In order to compute grain temperature for a model cloud of relatively
arbitrary geometry, one thus must  find an efficient method of
performing the angle integration of equation 2.  Once the grain
temperature is determined for all positions, a simple integral allows
one to derive the intensity distribution expected at a given
wavelength. One merely needs to choose the incident interstellar
radiation field $J_\nu^{\rm is}$ and the grain absorption efficiency
$Q_\nu$.

There are a number of choices which one can make concerning the grain
opacities. Mathis (1990) has tabulated values expected for ``standard''
interstellar grains and we have used these for comparison purposes.
However, within dense dust clouds, one expects grains to acquire ice
mantles and this will substantially change their optical properties. In
addition, one expects at the densities of interest to us that grain
coagulation may change the size distribution. We have therefore used
for many purposes the opacities computed by Ossenkopf \& Henning (1994,
hereafter OH)
for grains which have coagulated for a time of $10^5$ years.  As a
standard case, we have used their case for thin ice mantles and density
$10^6$ \percc \  but we note that in many situations, it is more
realistic for pre--protostellar cores to consider their ``thick ice
mantle case''. We have found however that for our purposes, the
difference between the two is not great as is the difference
between their solutions for different
densities. This is consistent with the fact 
that  Kramer et al. (1997)  find that the observed ratio of
infrared extinction to millimeter emissivity in IC5146  
(for visual extinctions in the range 20-30 magnitudes) is
compatible with the OH results.

In similar fashion, we have used as a comparison standard the solar
neighbourhood radiation field given by MMP.
Black (1994) has  updated  this work using results from the
FIRAS experiment on COBE  (Wright et al.  1991).  This suggests that
the MMP field is an underestimate in the mid and far IR . However, in
any case, the values used are an ``educated'' guess concerning the
field actually incident upon cores such as L1544 and one can expect
differences from one region to another. For example, it is clear that
small particle emission in the mid infrared can be an  important
component of the incident radiation field and one expects this to
depend on the far ultra--violet radiation from neighbouring O-B stars.
This can be expected to vary depending on how many massive stars have
formed in the vicinity of a given core.  One notes from the work of
Bacmann et al.~(2000, their table 1) that the intensity in the LW2 ISOCAM
filter (5-8.5\mic ) is  0.6 MJy/sr towards L1544 in Taurus
(a factor of order 3 larger than the average radiation field
at this wavelength according to Black (1994)) but 20.6
MJy/sr towards the OphD core (after correcting for zodiacal light).
This seems likely to be typical also of the small particle radiation at
 wavelengths up to 50\mic .

Finally, we need to define the density distribution of our models.  For
spherically symmetric cases, we have used the following density
distributions: ({\em i}\/) a homogeneous sphere, ({\em ii}\/) a
Bonnor-Ebert sphere (Bonnor~1956, Ebert~1955), and ({\em iii}\/) a
singular isothermal sphere (Chandrasekhar~1939, Shu~1977).  Case ({\em
i}\/) was chosen to test the accuracy of the numerical scheme, since in
this simple case exact solutions for the mean intensity and analytical
approximations for the dust temperature are easily derived.  Cases
({\em ii}\/) and ({\em iii}\/) represent equilibria of self-gravitating
non-magnetized isothermal clouds and have been extensively adopted to
model the density profiles of pre-protostellar cores. Notice that in
cases ({\em ii}\/) and ({\em iii}\/) the structure of the cloud is
computed assuming that the {\em gas} is isothermal. However, the {\em
dust} clearly is not. Moreover in the central parts of the core at densities
above $10^5$ \percc , the gas and dust temperatures are likely to
be coupled  (see e.g. Goldsmith 2001) and hence there is strictly
speaking an inconsistency which we will ignore in the following
discussion.  It is however a point which requires further study.

In all cases (including the non spherically symmetric models discussed
below), the model ``core'' was truncated at an outer radius $R_{\rm
out}$, and embedded in an envelope of radius $R_{\rm env}$ and uniform
density $\rho_{\rm env}$, equal to the density of the cloud at radius
$R_{\rm out}$. The rationale of this is that it simulates
the ``outer skin'' of the molecular cloud where a PDR
(Photon dominated region) shields the molecular gas against UV
and is consistent with the idea that the internal molecular cloud
pressure should not fall below the average interstellar pressure.
Thus, the
value of  $\rho_{\rm env}$  was chosen to simulate
the effect of the  internal GMC pressure $P_{G}$ confining the core.
One expects $P_{G}$ to be a factor of a few larger than the mean ISM
pressure of roughly $2\times 10^4$ K~\percc (see section 2 of McKee
1999). If we take $P_{G}$ equal to this minimum value and assume a
typical line width of 1 \kms \, we find an envelope density
corresponding to $n_{\rm H_{\rm 2}}$ = 1000 \percc.
 This outer PDR layer should provide shielding equivalent
to 1 magnitude of visual extinction and
we define $R_{\rm env}$ such that $n_{\rm H_{\rm 2}}\, (R_{\rm env}-R_{\rm
out})\,
= \, 1\times 10^{21}$ \cmsq .  We note that while these assumptions are
somewhat \adhoc, they have little influence on our results which concern
the high density core where the extinction is much higher than 1 magnitude.

Our non spherically symmetric models are based  on the results of CB
who have studied the development of infall driven by ambipolar
diffusion assuming axial symmetry.  They present results for a
``disk--like'' model and give the mid--plane density as a function of
radius at times $t_j$ ($j=0$,1,2,3...) when the central density
$n_{c}(t_{j})\, = \, 10^j\, n_{c0}$ with $n_{c0}=4.4\times 10^3$
\percc .  The times $t_{j}$ are 0, 2.27, 2.60, 2.66, 2.68 ... Myr from
which one sees that the last stages of the evolution are extremely
rapid. In this study, we consider the density distributions at times
$t_{2}$, $t_{3}$, and $t_{4}$ as being of most interest from the
standpoint of comparison with data for cores similar to L1544. We also
need to make an assumption about the density distribution in the
direction perpendicular to the disk midplane. We have assumed that
hydrostatic equilibrium is valid  in the $z$ direction (perpendicular
to the mid--plane) and hence that the density
can be expressed as $n(r,z)\, = \, n_{c}(t_{j})F_{CB}(r) \phi(z)$
with $F_{CB}(r)$ taken from the study of CB and $\phi (z)= \,
1/\cosh^2(z/H)$ where the scale height is $H=[kT/(2\pi{\rm
G}\rho)]^{1/2}$ and $\rho$ is the density.  For this purpose, the disk
is considered to be isothermal at a temperature of 12~K. Again,
this  is certainly not
self--consistent for the reasons mentioned earlier but this seems
a reasonable approximation to the density distribution of a CB
model.

\section{Analytical results for spherically symmetric clouds}
\label{analytical}

The results for spherically symmetric models are important to us both
because they can be treated  to some extent with analytic techniques
and because direct comparisons are possible with ERSM.  As a check on
our computational accuracy, we first show our results for the mean
intensity in a homogeneous spherical cloud compared with the analytic
solution.

\subsection{Mean intensity in a homogeneous sphere}

The solution of the full equation of radiative transfer in a homogeneous
sphere can be obtained analytically with the spherical harmonics method
(Flannery, Roberge, \& Rybicki~1980). In Appendix~A we show that
for a homogeneous cloud of radius $R$ and centre-to-edge optical depth
$\tau_{\nu c}$, the calculation of $J_\nu(r)$ from eq.~(\ref{Jrdef}) is straightforward
and can be expressed in terms of exponential-integral functions of the
argument $x_\pm=\tau_{\nu c}(1\pm x)$, where $x= r/R$:
\begin{eqnarray}
\lefteqn{J_\nu(x)=\frac{J^{\rm is}_\nu}{2(x_+-x_-)}\times} \nonumber \\
& & \left[x_+E_2(x_-)-x_-E_2(x_+)+x_-E_0(x_-)-x_+E_0(x_+)\right].
\label{Jexact}
\end{eqnarray}

Approximate expressions for the mean intensity can be obtained
in the limit of very large or very small $\tau_{\nu c}$.
In the optically thin limit eq.~(\ref{Jexact}) becomes
\be
J_\nu(x)\simeq J^{\rm is}_\nu\left\{1-\tau_{\nu c}\left[\frac{1}{2}
+\frac{1-x^2}{4x}\ln\left(\frac{1+x}{1-x}\right)\right]\right\},
\label{apprthin}
\ee
whereas an approximate formula (diverging for $x\rightarrow 1$)
for the optically thick case is
\be
J_\nu(x)\simeq J^{\rm is}_\nu
\frac{e^{-\tau_{\nu c}}}{2\tau_{\nu c} x}
\left(
 \frac{e^{ \tau_{\nu c}x}}{1-x}
-\frac{e^{-\tau_{\nu c}x}}{1+x}\right).
\label{apprthick}
\ee

\begin{figure}
\centerline{\psfig{file=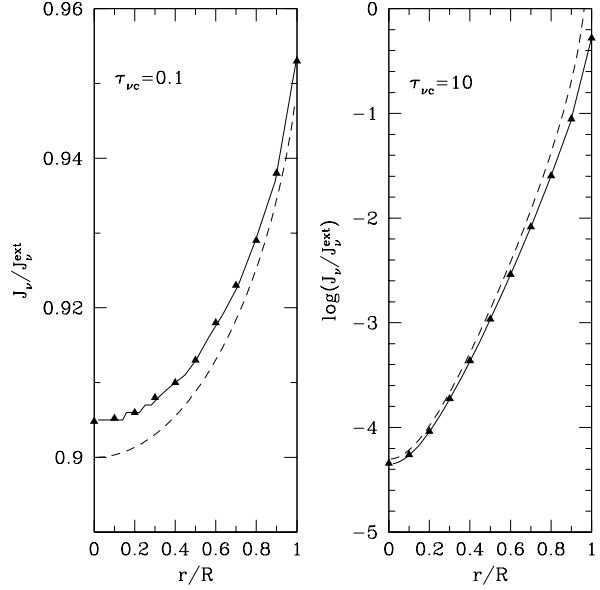,width=0.49\textwidth}}
\caption{Mean intensity $J_\nu$ for a homogeneous
spherical cloud as a function of radius $x=r/R$ computed by numerical
integration of eq.~(\ref{Jrdef}) ({\em solid lines}) compared with the
analytic
formula eq.~(\ref{Jexact}) ({\em triangles}). The approximated expressions
(\ref{apprthin}) and (\ref{apprthick}) for the optically thin and thick
cases are shown by {\em dashed lines}.  Results are shown for two
model clouds with $\tau_{\nu c}=0.1$ and 10.}
\label{homsphere}
\end{figure}

In Figure~\ref{homsphere}, we compare the exact and approximate
analytical results with that obtained numerically integrating
eq.~(\ref{Jrdef}). We see that the numerical code reproduces the
analytic result very well.  Thus, in simple geometries, the procedure
used for the angle integration is adequate.

\subsection{The dust temperature at the centre of a spherical cloud}

There are great advantages in developing analytic formulae which  allow
an approximate estimate of the temperature in dense pre--protostellar
cores.  This permits for example a discussion  of scaling relationships
for different external radiation fields and opacities. We have
accordingly attempted to derive an analytical estimate of the
temperature at the center of a spherically symmetric cloud for
comparison with our numerical results.

In order to develop an analytic approximation, we must first make some
crude approximations concerning the dust opacity and external radiation
field in cores such as L1544.  Following Black~(1994) and MMP,
the interstellar radiation field at
wavelengths above 0.1~$\mu$m can be represented by the sum of four
contributions: (1) an optical-NIR component peaking at $\lambda_{\rm p}=
1$~$\mu$m, due to the emission of disk dwarf and giant stars; (2) the
diffuse FIR emission from dust grains, peaking at $\lambda_{\rm p} =
100$~$\mu$m; (3) mid--IR radiation from small non--thermally heated
grains in the range 5--100~\mic \ and (4)
the cosmic background radiation, peaking at $\lambda_{\rm p}
=1$~mm. Components 1,2, and 4
can be represented by a single (or a sum
of) modified black-bodies at temperatures $T_i$ of the kind
\be
J_\nu^{\rm is}=\frac{2h\nu^3}{c^2}\left(\frac{\lambda_{\rm p}}{\lambda}\right)^p
\sum_i\frac{W_i}{e^{h\nu/kT_i}-1},
\label{rfield}
\ee
where $\lambda_{\rm p}$ is the peak wavelength and $W_i$ are dilution
factors. The values of the parameters $\lambda_{\rm p}$, $W_i$ and
$T_i$ are listed in Table~\ref{isf_param}.  The third (MIR) component
is found to be variable near dense cores such as L1544 and can be
approximated as a power law. In the following discussion however, we at
first neglect it and then make an estimate of its importance.

As in the numerical computations, we have based our  opacity law on the
work of OH (thin ice, density $10^6$ \percc
).  For the purposes of our analytic work, a very crude piecewise power
law fit with breaks at 10~\mic\ and 400~\mic\ has been used (see
Table~\ref{op_param}). We then assume that in a sufficiently large range of
wavelengths around the peak of each component, the dust absorption
efficiency $Q_\nu$ can be approximated by a power-law, $Q_\nu \propto
(\lambda_{\rm p}/\lambda)^\alpha$.  The values of the parameters
$\lambda_{\rm p}$ and $\alpha$ used by us are listed in
Table~\ref{op_param}.

Let the radiation absorbed (indicated by the subscript ``{\it a}'') by
dust be concentrated around a frequency $\nu_a$, and the radiation emitted
(indicated by the subscript ``{\it e}'') by dust be characterized by a
frequency $\nu_e$.  We take $\nu_e$ and $\alpha_e$ equal to the
values for absorption in the FIR (see Table~\ref{op_param}), and we
consider the heating of the dust by each component of the radiation
field separately.

With these assumptions, the mean intensity at the cloud's center is
\be
J_\nu(0)=J_\nu^{\rm is}\exp{(-\tau_{\nu c})}=
J_\nu^{\rm is}\exp{\left[-\tau_a\left(\frac{\nu}{\nu_a}\right)^{\alpha_a}\right]},
\ee
where $\tau_a$ is the centre-to-edge optical depth of the cloud at frequency $\nu_a$.
Then, defining $\beta_d\equiv kT_d/h\nu_e$, $\beta_i\equiv
kT_i/h\nu_a$, eq.~(\ref{therm_eq}) becomes
\be
{\cal A}=\frac{Q_a}{Q_e}\left(\frac{\nu_a}{\nu_e}\right)^4\left(\frac{\nu_a}{\nu_{\rm p}}\right)^p
\sum_i W_i {\cal B}_i,
\ee
where $Q_a$ and $Q_e$ are the values of $Q_\nu$ at frequencies $\nu_a$ and $\nu_e$,
respectively, and
\be
{\cal A}=\int_0^\infty\frac{t^{\alpha_e+3}}{e^{t/\beta_d}-1}\;dt
=\Gamma(\alpha_e+4)\zeta(\alpha_e+4)\beta_d^{\alpha_e+4},
\ee
\be
{\cal B}_i= \int_0^\infty\frac{t^q}
{e^{t/\beta_i}-1}\exp{(-\tau_a t^{\alpha_a})}\;dt,
\ee
with $q=\alpha_a+p+3$. In the above, $\Gamma$ and $\zeta$ are the gamma
and Riemann zeta functions (see Abramovitz \& Stegun~1965).  The
integral in the equation for ${\cal B}_i$ can be easily evaluated in
the two limiting situations $\tau_a\ll 1$ and $\tau_a\gg 1$.  For
$\tau_a\ll 1$ the exponential in ${\cal B}_i$ can be expanded as a
Taylor series, giving, to first order in $\tau_a$,
\begin{eqnarray}
\lefteqn{{\cal B}_i\simeq \Gamma(q+1)\zeta(q+1)\beta_i^{q+1}} \nonumber \\
& & -\tau_a\Gamma(q+\alpha_a+1) \zeta(q+\alpha_a+1)\beta_i^{q+\alpha_a+1}+\ldots.
\label{approx_thin}
\end{eqnarray}

Thus, the dust temperature at the center is given, to lowest order, by
\be
\frac{k T_d}{h\nu_e}\simeq C
\left(\sum_i W_i\beta_i^{q+1}\right)^\frac{1}{\alpha_e+4},
\label{tcent_thin}
\ee
where
\be
C=\left[\frac{\Gamma(q+1)\zeta(q+1)}{\Gamma(\alpha_e+4)\zeta(\alpha_e+4)}
\frac{Q_a}{Q_e}\left(\frac{\nu_a}{\nu_e}\right)^4
\left(\frac{\nu_a}{\nu_p}\right)^p\right]^\frac{1}{\alpha_e+4}.
\ee

We note here that for the case of interest in which the dust is heated
by (optically thin) FIR radiation, $Q_a=Q_e$, $\nu_a=\nu_p=\nu_e$ and one finds that
$C\simeq 1.68$ is a constant slightly dependent on the power law index
of the dust opacity but otherwise independent of dust characteristics.

For $\tau_a\gg 1$ the integrand in ${\cal B}_i$ can be expanded
as a Taylor series for small $t$, obtaining the approximate result
\begin{eqnarray}
\lefteqn{{\cal B}_i\simeq
\frac{\beta_i}{\alpha_a}\Gamma\left(\frac{q}{\alpha_a}\right)
\tau_a^{-q/\alpha_a}} \nonumber \\
& & - \frac{1}{2\alpha_a}\Gamma\left(\frac{q+1}{\alpha_a}\right)
\tau_0^{-(q+1)/\alpha_a}+\ldots.
\label{approx_thick}
\end{eqnarray}
Thus, the dust temperature at the center is given, to lowest order, by
\be
\frac{k T_d}{h\nu_e}\simeq D\tau_a^{-q/[(\alpha_e+4)\alpha_a]}
\left(\sum_i W_i\beta_i\right)^\frac{1}{\alpha_e+4},
\label{tcent_thick}
\ee
where
\be
D=\left[\frac{\Gamma(\frac{q}{\alpha_a})}{\alpha_a
\Gamma(\alpha_e+4)\zeta(\alpha_e+4)}
\frac{Q_a}{Q_e}\left(\frac{\nu_a}{\nu_e}\right)^4
\left(\frac{\nu_a}{\nu_p}\right)^p\right]^\frac{1}{\alpha_e+4}.
\ee
Notice that the dependence of the dust temperature on the intensity of
the external radiation is rather weak, $T_d\propto
W^{1/(\alpha_e+4)}$. Again, this expression simplifies considerably
when the dust is heated by (optically thick) FIR radiation,
and $D\simeq 0.59$ in this case.

The above discussion has neglected the contribution of the
MIR radiation. We approximate the external MIR field with
a power law spectrum
in the wavelength range 10--100~\mic \ and with a lower frequency
cut--off at 100\mic ,
\be
J_\nu^{\rm is}=W\frac{2h \nu_p^3}{c^2}
\left(\frac{\nu}{\nu_p}\right)^p,
\ee
with the values of $W$, $\lambda_p$ and $p$ given in Table~\ref{isf_param}.
Notice that $p<0$ in this case and thus for the ``MIR'' radiation,
there is approximate cancellation in the integral on the
right hand side of equation ~\ref{therm_eq} between the
frequency dependence of $Q_{\nu}$ and that of the external MIR
field. The consequence is that the main contribution to the
integral comes from frequencies where the optical depth is of order unity
(i.e. at a wavelength of roughly
 20\mic \ for $A_{V}$=30 mag.).

Substituting the above expression for the MIR field
in eq.~(\ref{therm_eq}), we obtain
\be
{\cal A}=W\left(\frac{\nu_p}{\nu_a}\right)^4\frac{\Gamma(s,\tau_a)}{\alpha_a}
\tau_a^{-s},
\label{tcent_mir}
\ee
where $s=(\alpha_a+p+1)/\alpha_a\, = \, 0.5$ for our parameters,
and $\Gamma(s,\tau_a)$ is the incomplete
gamma function. Here $\tau_a$ is the cloud's optical depth at 100~\mic,
which becomes unity at $A_V\simeq 414$. Below this value, the incomplete
gamma function can be expanded for small $\tau_a$, obtaining
\be
\Gamma(s,\tau_a)\simeq \Gamma(s)+\frac{\tau_a^s}{s}-\frac{\tau_a^{s+1}}{s+1}+\ldots.
\label{gamma_exp}
\ee

In eq.~(\ref{tcent_thin}), (\ref{tcent_thick}), (\ref{tcent_mir}) and
(\ref{gamma_exp}), it is convenient to express the optical depth
$\tau_a$ in terms of the optical extinction $A_V$,
\be
A_V=1.086\tau_V=1.086\left(\frac{Q_V}{Q_a}\right)\tau_a,
\ee
where the ratio in parentheses can be computed for each components with
the parameters listed in Table~\ref{op_param} for the Ossenkopf \&
Henning~(1994) opacity.  We see that a cloud is optically thin to V-NIR
radiation for $A_V\ll 3$, to FIR radiation for $A_V\ll 700$, and to the
cosmic background radiation for $A_V\ll 6\times 10^4$.  Thus, with
Black~(1994) standard interstellar field, the central temperature in a
``diffuse'' cloud core of typical centre-to-edge extinction $A_V\simeq 1$--2
is determined by optically thin absorption of V-NIR and MIR radiation (the
effects of an external UV field are not considered here) whereas for
$10\simless A_V \simless 400$ is determined by optically thick absorption of V-NIR
radiation and optically thin absorption of MIR and FIR radiation.

Inserting the numerical
values given in Appendix~B (and keeping terms up to second order in
$\tau_a$), the formulae above give
\be
T_d(\mbox{V-NIR, thin}) \simeq 16.5-3.6A_V+\ldots,
\label{vnirthin}
\ee
\be
T_d(\mbox{V-NIR, thick}) \simeq 43 A_V^{-0.56}-77 A_V^{-1.28}+\ldots,
\label{vnirthick}
\ee
\be
T_d(\mbox{FIR, thin}) \simeq 6.2-0.0031 A_V+\ldots,
\label{firthin}
\ee
\begin{eqnarray}
\lefteqn{T_d(\mbox{MIR, thin}) \simeq 7.9A_V^{-0.089}} \nonumber \\
& & \times(1.8-0.098A_V^{0.5}+7.9\times 10^{-5}A_V^{1.5}+\ldots)^{1/5.6},
\label{mirthin}
\end{eqnarray}
where $T_d$ is in $K$.

Combining eq.~(\ref{vnirthin})--(\ref{mirthin}), one can obtain
simple analytical expressions for the dust temperature (in $K$) at the
cloud's centre,
\be
T_d\simeq [T_d(\mbox{V-NIR, thin})^{5.6}
+T_d(\mbox{MIR, thin})^{5.6}]^{1/5.6},
\label{tdlow}
\ee
for $A_V\simeq 1$--2, and
\begin{eqnarray}
\lefteqn{T_d\simeq [T_d(\mbox{V-NIR, thick})^{5.6}
+T_d(\mbox{MIR, thin})^{5.6}} \nonumber \\
& & +T_d(\mbox{FIR, thin})^{5.6}]^{1/5.6},
\label{tdhigh}
\end{eqnarray}
for $10 \simless A_V\simless 400$. Notice that increasing the intensity
of the radiation field by a factor $g$ results in an increase of the
dust temperature by a factor $g^{1/5.6}$.

\begin{figure}
\centerline{\psfig{file=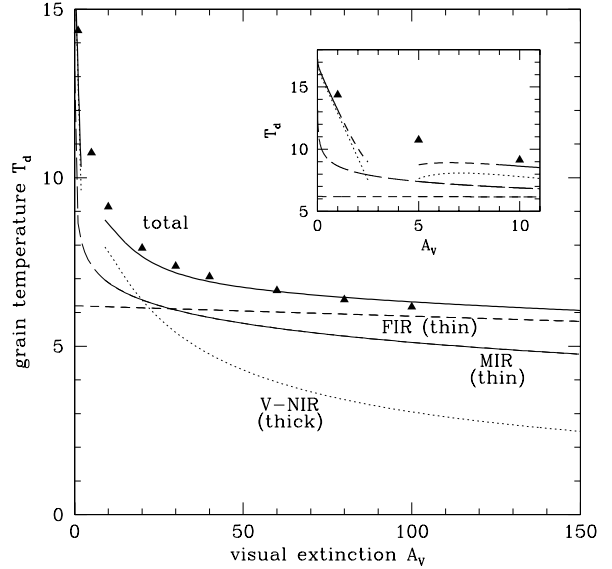,width=0.49\textwidth}}
\caption{Dust temperature $T_{\rm d}$ at the centre of spherical
clouds of visual extinction $A_V$. The contribution of
V-NIR radiation ({\it dotted line}\/), MIR radiation ({\it long-dashed
line)}, FIR radiation ({\it short-dashed line}\/), and the total
({\it thick solid line}\/), obtained with the analytical formulae
of Sect.~3.2 are compared with the numerical results ({\it triangles}\/).
The inset shows a blow--up of the figure for extinctions less than
11 magnitudes.}
\label{analytic_temp}
\end{figure}

We have checked these approximate formulae against the numerical
computations and the results are shown in Figure~\ref{analytic_temp}.
Here we compare in the first place the numerical model predictions for
the temperature at the center of a Bonnor--Ebert sphere with that
derived from eq.~(\ref{tdlow}) and (\ref{tdhigh}).
We see that the analytic formula  is in
agreement with the more accurate numerical results within $\sim 1$~K .

\begin{figure}
\centerline{\psfig{file=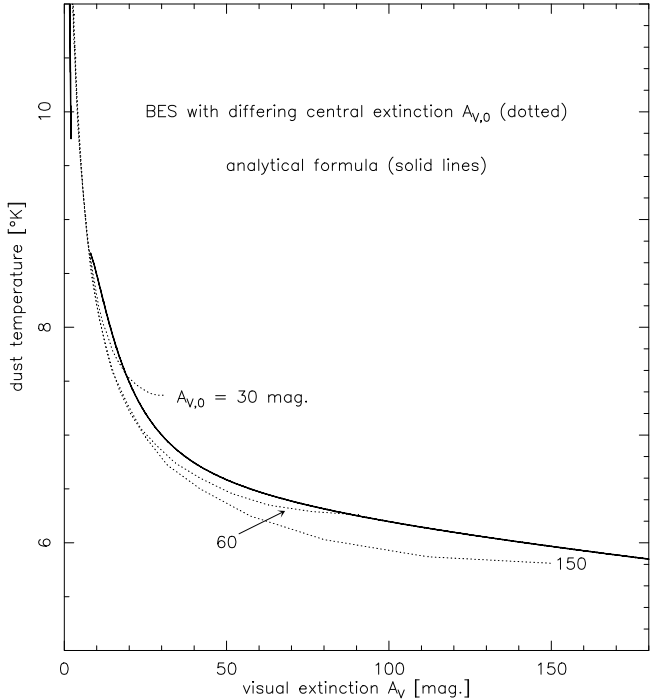,width=0.49\textwidth}}
  \caption
  { Comparison of the radial dependence of
  dust temperature computed for model Bonnor--Ebert
  spheres with central extinctions 30, 90, and 150 magnitudes
    ({\it dotted curves}) with results from analytic
  approximations ({\it thick full lines}) in eq.~(\ref{tdhigh}) and (\ref{tdlow}).
  Here, the extinction $A_V$ in the analytic formulae
  has been interpreted as the minimum extinction along a ray
  emerging from the core.
    }
  \label{tanalytic}
\end{figure}

It is also of interest to compare the expectations from the analytic
formula with that calculated as a function of radius in a given
Bonnor-Ebert sphere. Here, we merely use the ``intuitive
approximation'' that the appropriate optical depth to substitute in eq.~(\ref{tdhigh})
is the minimum optical depth along a ray to the exterior of the
core. We see in Fig. ~\ref{tanalytic} that this approximation is also quite
reasonable suggesting that, in a number of situations, one
can obtain an estimate of the dust temperature as a function of
position using eq. (\ref{tdhigh}). Obviously however, this must
be used with caution. Nevertheless, it should as a rule
be more accurate
than the isothermal assumption.

\section{Numerical computations of the  dust temperature distribution}

In this section, we present our results obtained numerically integrating
eq.~(\ref{therm_eq}) and (\ref{Jrdef}).

\subsection{SIS and Bonnor-Ebert spheres}

\begin{figure}
\centerline{\psfig{file=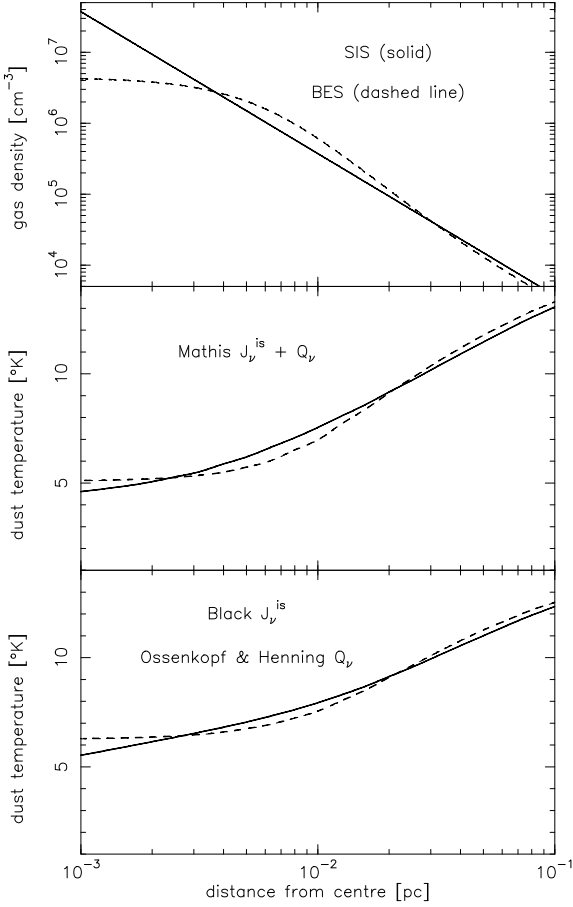,width=0.49\textwidth}}
\caption{Dust temperature as function of radius computed by the numerical
model for the SIS ({\it full lines}) and Bonnor-Ebert sphere
(central density $4.4 \times 10^{6}$ \percc , {\it dashes}) distributions.
The top panel shows the density distribution and the bottom two panels
show the temperature distribution we compute for two different
choices of external radiation field and grain optical parameters.
The center panel shows results for the MMP radiation
field together with the grain opacities recommended by Mathis~(1990).
The bottom panel shows results obtained using the opacities of
OH (their col. 5) and the radiation field of
Black (1994).}
\label{sisbeb}
\end{figure}

We first consider inhomogeneous spherically symmetric density
distributions. Two obvious cases of interest are the singular
isothermal sphere (SIS) solution and the Bonnor-Ebert sphere (BES). In
Fig.~\ref{sisbeb}, we show the results we have obtained for these density
distributions. Both models are obtained from the equation of hydrostatic
equilibrium assuming a gas temperature of 12 K.
For the BES, we have chosen a model with a
central gas density $n_{\rm H_{\rm 2}}$ = $4.4\times 10^{6}$ \percc
\ similar to some observed pre--protostellar cores. One
notes first
that the two density distributions only differ greatly at small radii
(less than $10^4$~AU or 0.03~pc) where the BES distribution ``flattens
out''. The computed temperature distributions for the two cases are
close to being identical even in the central regions. There is in both
cases a substantial fall off from values above 15~K at the edge (in the
constant density envelope mentioned above) to values as low as 5-6~K at
a radius of $10^{-3}$ pc or 200~AU.

These results do not change greatly for different choices of grain
opacity and radiation field.  However, a larger radiation field,
as estimated by Black~(1994) in the range 10--50 \mic , causes a slight
increase in the central dust temperature.  In general, we conclude that
in structures such as these (with visual extinctions of order 100
mag.), the central temperature can be expected to drop by roughly a
factor of 2--3 relative to that measured on  scales of $\sim $ 0.1 pc.
We have
compared our results with those of ERSM for similar parameters and find
that they obtain slightly higher temperatures (7.5~K as compared to
7.2~K for the same conditions) which may be due to our neglect of
heating due to the radiation of the core itself. However, the general
behavior seems very similar for both codes although, as noted above,
our neglect of heating by core radiation is not valid for the SIS 
distribution at small radii.

\subsection{Temperature distribution in the CB model}

\begin{figure}
\centerline{\psfig{file=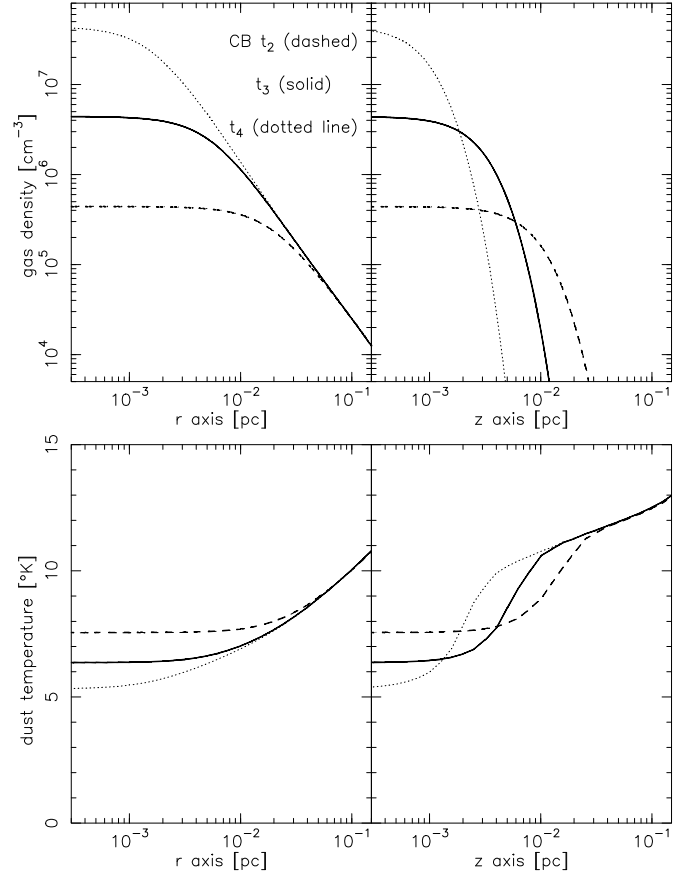,width=0.49\textwidth}}
\caption {Dust temperature variation with position in the models of CB.
We show in the top panels the density distribution implied by their
model in the midplane of the disk (top left) as a function of radius
$r$ and  perpendicular to the disk as a function of height above the
midplane $z$ (top right). Results are shown for three different times
in the evolution: at 2.60~Myr ($t_2$) when the central ${\rm H_{\rm 2}}$
density is
$4.4\times 10^5$ \percc, at 2.66
Myr
($t_3$) when the central density is $4.4\times 10^6$ \percc , and at 2.68
Myr when the central density is $4.4\times 10^7$ \percc . The bottom
panels
show the corresponding temperature distributions computed by our code
as a function of $r$ and $z$. Dashes are for time $t_2$, the full
line for $t_3$, and dots for $t_4$.}
\label{tdist_cb}
\end{figure}

We now consider the temperature distribution to be expected within the
model  proposed by CB for L1544.  As discussed in section 2, this is a
disk--like axisymmetric model. In this discussion, we consider the
radial distribution of dust temperature in the mid--plane of the disk
as well as the temperature distribution along the axis of symmetry
perpendicular to the disk. The calculations are presented for the
density distributions corresponding to three evolutionary times $t_2 =
2.60$~Myr, $t_3 = 2.66$~Myr, and $t_4 = 2.68$~Myr during which time the
central density evolves from a molecular hydrogen density of $4.4\times
10^5$ \percc \ to $4.4\times 10^7$ \percc (see section 2).

In Fig.~\ref{tdist_cb} we present results for these models. One notes
the fact that the CB models are relatively ``thick'' disks in that
their extent along the $z$-axis is an appreciable fraction of their
extent in $r$. The dust temperature computed by our models behaves in
similar fashion to that obtained in our spherically symmetric models.
The main effect of the disk structure is simply that one finds dust at
temperatures of 10~K or higher much closer to the center of the core
than in the spherically symmetric case.

\begin{figure}
\centerline{\psfig{file=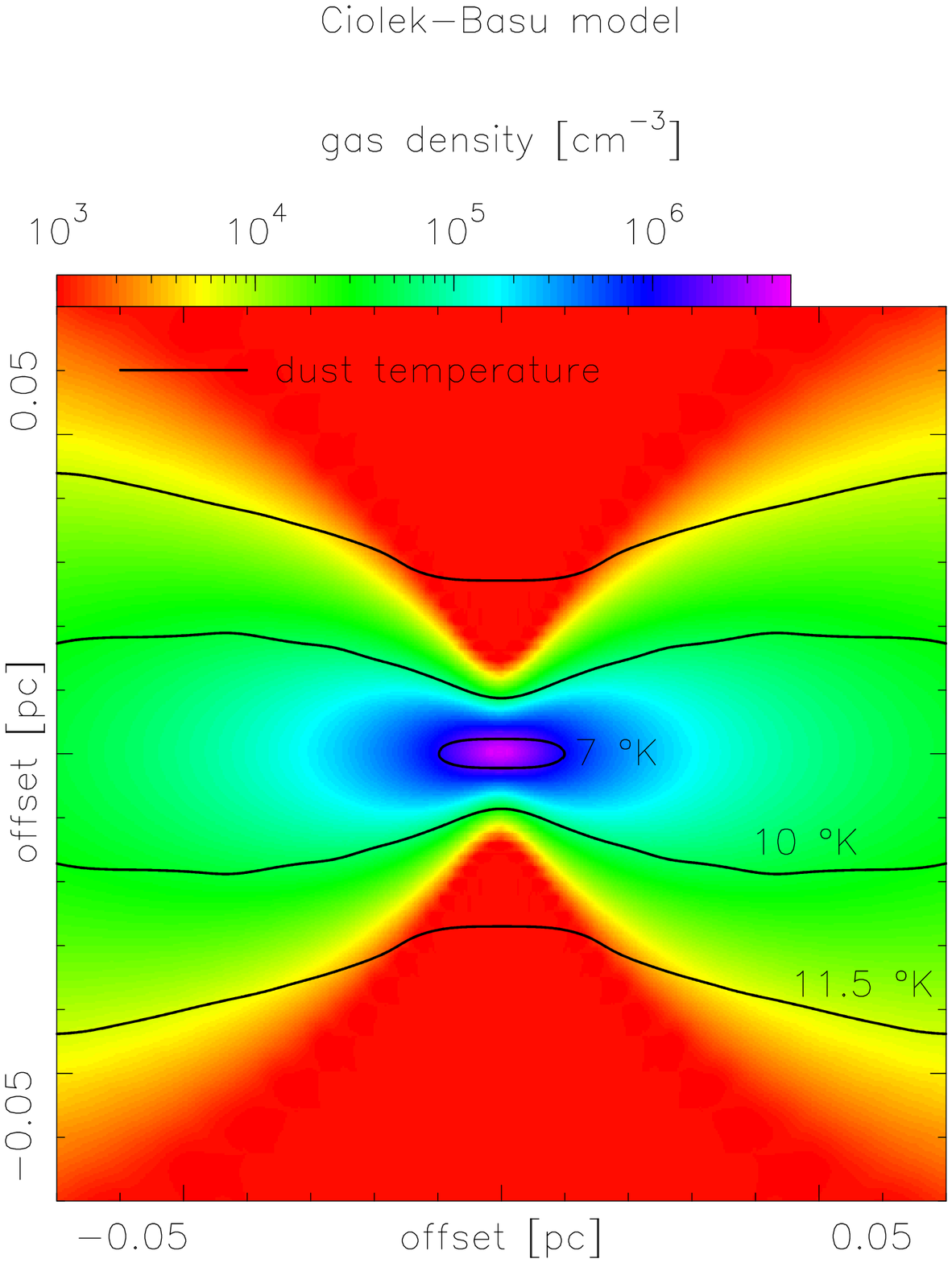,width=0.49\textwidth}}
\caption {Dust density and temperature variation in the model
of CB at time $t_3$ (2.66 Myr with central density $4.4\times 10^6$
\percc )seen in the plane perpendicular to the
disk midplane. The density variation shown in color is that
of a flared disk. The temperature variation is given by the
contours.}
\label{tdcb_edge}
\end{figure}

Another view of the temperature variation is presented in
Fig.~\ref{tdcb_edge} which shows a cut perpendicular to the
disk midplane. One sees that close to the midplane, the attenuation
of the external radiation field is large and the temperature
remains relatively low while in the perpendicular direction, it
rises more rapidly. The temperature gradient is larger along the
minor axis if one examines this model edge on. In practise, one
must take into account effects due to finite inclination as well as the
contributions from foreground and background layers.

\section{Comparison of Models with Data}

In this section, we present calculations of the mm--submm
emission predicted by the models discussed above.
We compute the expected intensity distributions
and compare with observations of L1544.  We
consider first the spherically symmetric cases.

\subsection{Intensity Distribution for spherically symmetric models}

\begin{figure}
\centerline{\psfig{file=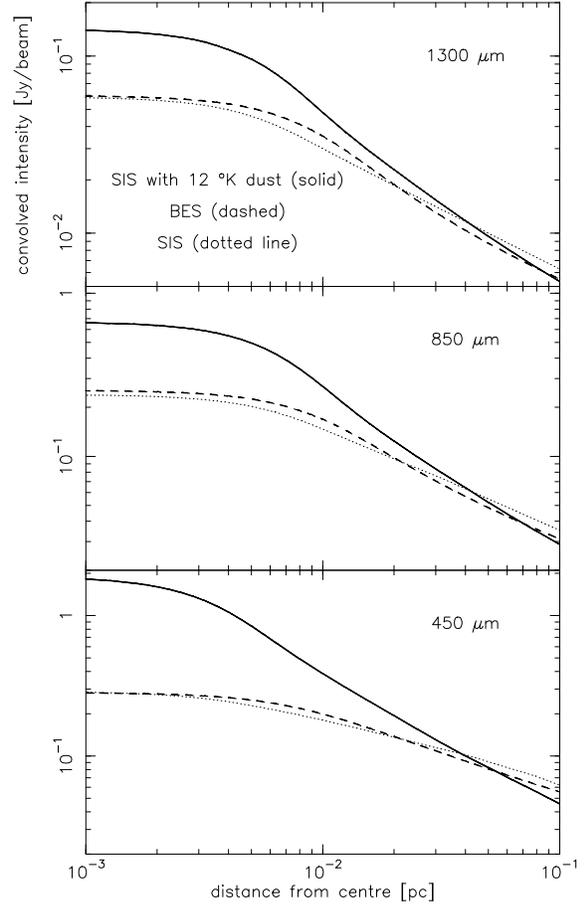,width=0.49\textwidth}}
\caption {Intensity as a function of  projected radius for the SIS and BES
models
discussed in section 2  and shown in Fig.~\ref{sisbeb}.  The top panel
shows results at
a wavelength of 1.3~mm, the center panel at a wavelength of 850~\mic
\ and the bottom panel at a wavelength of 450 \mic .  We show results
for the Bonnor--Ebert sphere ({\it dashed}) and SIS models ({\it dotted
line}) with the  Black radiation field and OH opacities.  For comparison,
we also show the expected distributions for a SIS model at a dust
temperature of 12~K
({\it solid}). }
\label{idist_ss}
\end{figure}

Using the dust temperature distribution discussed earlier, we have
computed the expected intensity distributions at three wavelengths for
which observational data are available in the literature. These results
have been obtained simply integrating the transport equation (with
our calculated temperature distribution) across our model core.
We have convolved the resultant intensity distribution with a
``gaussian beam'' of 13, 15 and 8\arcsec (at 140 pc) for 1.3 mm, 850, and
450 \mic , respectively.
We present in Fig.~\ref{idist_ss} results using the Black (1994) radiation 
field and
compare the expected intensity profile at wavelengths mentioned above from
the BES and SIS models.

One sees that at all wavelengths but especially at 450 \mic , the
profile is much flatter using our calculated temperature distribution
than in the isothermal case. This just reflects the fact that the cold
core nucleus predicted by our calculations does not contribute
substantially at the shorter wavelengths.  Even at 1300~\mic , the
contrast between the intensity at  offset $10^{-3}$ pc 
(200 AU) and 0.1 pc is
26 in the isothermal case and 9
using our computed temperature gradient.  Thus we conclude (in
agreement with ERSM) that conclusions about the density structure based
upon the intensity distribution of the dust continuum emission must
take temperature gradients into account.

Another clear result from Fig.~\ref{idist_ss} is that the predicted
intensity distributions for the BES and SIS models are essentially
identical. The low temperature of the central nucleus causes the
observed intensity profile to be insensitive to the density
distribution inside 0.01 pc (2000 AU).  This means that one should be
very cautious about claims for a flat central density distribution
based upon maps of millimeter dust emission. On the other hand,
derivations of the density profile based upon either NIR reddening
(Alves et al. 2001) or mid-IR extinction (Bacmann et al. 2000) are
insensitive to dust temperature and hence are to be preferred for such
purposes.

\subsection{Intensity Distribution for the CB model}

\begin{figure}
\centerline{\psfig{file=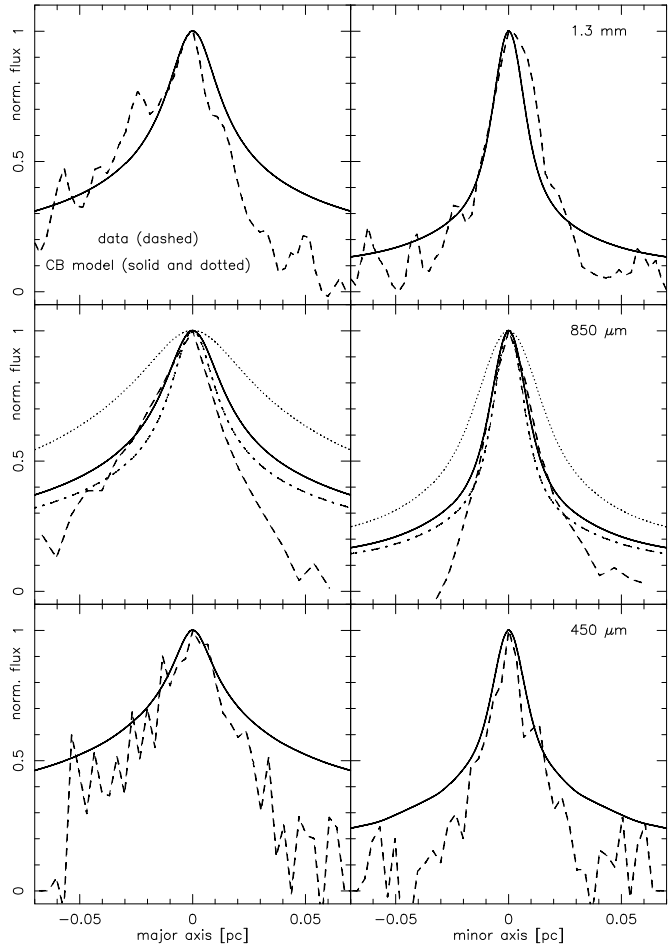,width=0.49\textwidth}}
\caption {Intensity distribution ({\it full lines})
for wavelengths of 1.3~mm (top), 850~\mic
\ (center), and 450~\mic \ (bottom) predicted by the CB model along the
major axis (left hand panels) of L1544  and along the minor axis (right
hand panels) assuming the density distribution  at time $t_{3}$
(2.66 Myr) with central density $4.4\times 10^6$ \percc \ . 
In the panels for 850~\mic \ we also show profiles computed for the CB
model
at time $t_{4}$ ({\it dash--dot}, 2.68 Myr) and $t_{2}$ ({\it dots}, 2.60
Myr) (see caption for Fig.
~\ref{tdist_cb}). An inclination of
$16^{\circ}$ to
the plane of the sky has been assumed.  For comparison, we show also
observed cuts along the major and minor axes of L1544 taken from
Ward-Thompson et al. (1999) (1.3~mm data) and Shirley et al. (2000) (850\mic
\ and 450~\mic \ data). The model intensity distribution has been convolved
with gaussian beams with an HPBW of 13, 15 and 8 arc seconds
for 1.3~mm, 850~\mic \
and 450\mic, respectively, i.e. the same observational beam as the data.}
\label{int_cb}
\end{figure}

The spherically symmetric models discussed above clearly do not apply
to cores such as L1544 which has an extremely elongated intensity
distribution when mapped at millimeter wavelengths (as well as an
elongated appearance when observed in absorption at 7~\mic \ , Bacmann
et al. 2000).  We therefore now consider the expected intensity
distributions for the models of CB shown in Fig.~\ref{tdist_cb}.
One should note that these were originally
constructed to fit the observed millimeter
continuum intensity distribution for L1544 assuming  isothermal dust.
We now consider the consequences of the more realistic temperature
distribution discussed above  and compare with the available
observations. We have again here used dust opacities from OH
(thin ice, col. 5) and the external radiation field of Black~(1994).

In Fig.~\ref{int_cb}, we show the resulting intensity distributions for
the epoch $t_3$ of CB (central density $4.4\times 10^6$ \percc ).
We have assumed an inclination of
$16^{\circ}$ with respect to the plane of the sky as suggested by CB and
compare with cuts along the major
and minor axes from the work of Ward-Thompson et al. (1999) and Shirley et
al. (2000). We conclude that despite departures from isothermality,
the agreement between model and observed
intensity distribution is reasonable at all three frequencies for time
$t_3$. It certainly gives a better fit than the profile for time $t_2$
although it would be difficult to distinguish between $t_3$ (2.66 Myr) and
$t_4$ (2.68 Myr) on this basis. Thus we confirm the result of CB that this
stage in the development
of their model fits the observed intensity distribution.
We note nonetheless that in all cuts along the major axis, the fall off
in observed intensity at positive offsets is more rapid than in the
models.

\begin{figure}
\centerline{\psfig{file=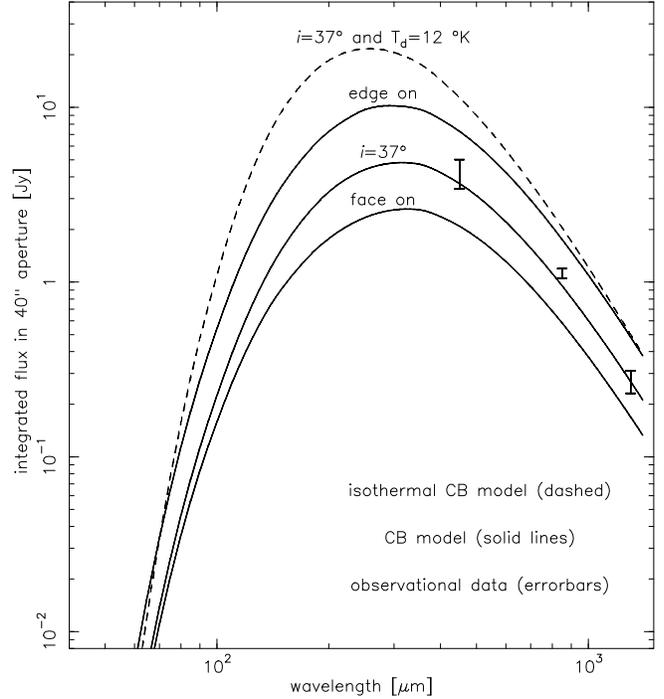,width=0.49\textwidth}}
\caption
{Observed spectral energy distribution for L1544 compared with model
predictions ({\it solid lines}) for the flux in a 40\arcsec \ aperture as
a
function of wavelength.  The model results are for the flux (Jy) in an
aperture of 40\arcsec \ HPBW towards the peak of the L1544 dust
emission. Observational data (error bars) are from Shirley et al.
(2000) and  Ward--Thompson et al.~(1999).  The model data are for 3
different inclinations of a CB disk (time $t_{3}$)
to the plane of the sky. The dashed line shows for comparison the
expected SED for an isothermal CB disk inclined at 37$^{\circ}$.}
\label{modelsed}
\end{figure}

Another characteristic that a model should fit is the observed
spectral energy distribution or SED. A complication here
is that observations typically resolve pre--protostellar
cores and that the SED thus depends sensitively on the assumed aperture.
In Fig.~\ref{modelsed}, we
present the expected spectrum for L1544 compared with observations
from Shirley et al. (2000) for an assumed 40\arcsec \ aperture
(a compromise value). One sees that edge--on disks of this type are
considerably stronger and easier to observe than face--on disks.
This is because
(even with a 40~\arcsec beam at a distance of 140 pc)
we resolve the source and as at these wavelengths, the model is
optically thin, the flux is greater for the edge--on case where
the column density is larger. We show for comparison the SED  for the
same model density distribution assumed to be isothermal and with an
inclination of 37$^{\circ}$. The isothermal model is slightly
``hotter'' than the model with computed temperature distribution
and has a somewhat steeper long wavelength spectrum.

 We get a reasonable agreement with
the observed SED for an inclination of $37^{\circ}$  though this result
reflects the assumed dust opacity and should  be treated
with caution. There is still considerable uncertainty concerning the
nature of the ice mantles on the dust grains at the heart of
pre--protostellar cores like L1544. We conclude nevertheless
that the non--isothermal
CB model at time $t_3$ does indeed give a reasonable fit to the
observed flux distribution.
The effects of departures from isothermality on the SED are probably small
in comparison to the observational difficulties
(e.g. due to pointing errors) of obtaining  matched apertures at
different frequencies towards an extended source.
This is both because the fluxes from a 40\arcsec \ region
are not very sensitive to the
central density distribution (i.e. the central 2000 AU) and because the
low
temperature causes a high density nucleus (if present) to have much
less weight. This is in particular true at 850~\mic \ and higher
frequencies and from this point of view, 1.3~mm observations of high
angular resolution are needed to probe the nucleus of such cores.


\begin{figure}
\centerline{\psfig{file=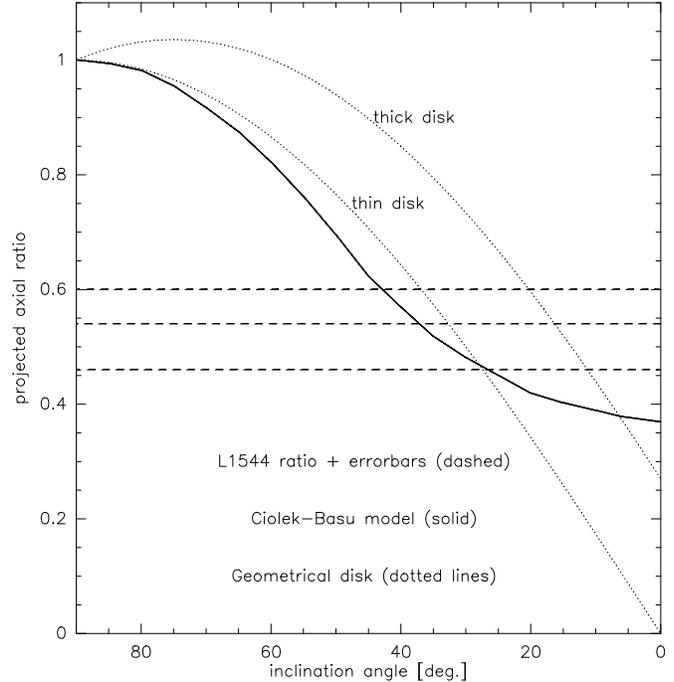,width=0.49\textwidth}}
\caption
{Ratio of half-power sizes along the major and minor
axes computed for CB model $t_{3}$ as a function of
inclination of the mid--plane relative to
the line of sight (full line).
The measured ratio for L1544 is shown for comparison
({\it dashed horizontal line}) with upper and lower limits.
The dotted lines show the axis ratios expected for a geometrically
thin disk ($\sin {i}$) and for a geometrically thick disk
using eq. (1) from CB ({\it upper dotted curve})
with  ratio of disk thickness to radius 0.27.}
\label{hpratio}
\end{figure}

 As seen above, if we interpret cores like L1544 in terms of the
 CB model, the inclination of the CB disk to the line of sight
 is a critical parameter. It is also important for the
 interpretation of
 Zeeman effect observations (Crutcher \& Troland 2001).
This suggests that, for a model
like that of CB, it is useful to estimate the inclination based upon
the ratio of major and minor axes in a tracer such as
the millimeter continuum.  This must be done using so far as possible the
true intensity distribution of the model core.
With this in mind, we have attempted to calibrate the measurement
using CB's model at time $t_{3}$ and
show the results in Fig.~\ref{hpratio}.  We compare the
results from the numerical computations with the formula
given by~CB for a disk of finite thickness
as well as the simple result for a thin
disk. Our computed axis ratio behaves rather  like a thin
disk at moderate inclinations but is close to the expected
result in the edge--on case. We obtain from the
axis ratios
a best fit for an inclination of $37^{\circ}$ consistent with the SED
of Fig. ~\ref{modelsed} shown above.
 However, our data would certainly accomodate an inclination
 of 25$^{\circ}$ and (as Fig.~\ref{int_cb} demonstrates) we cannot exclude
16$^{\circ}$.

\section{Discussion and Conclusions}

This study has shown that departures from isothermality in
pre--protostellar cores, while not a dominant effect, should be taken
into account when considering  core structure at times just prior to
the onset of collapse. This is of importance when attempting to deduce
the density structure of such regions based on maps of dust emission.
It also  may be important when considering the stability of such
cores. Above densities of $10^5$ \percc \ , one expects gas and dust
temperatures to be coupled (see Goldsmith 2001, Kr\"{u}gel \& Walmsley
1984).  Thus at higher densities in the center of cores such as L1544,
one can expect the gas temperature to follow the dust temperature.  A
decrease in dust temperature in the nucleus of such cores requires a
steeper density gradient than in an isothermal model and may have some
consequences for the behavior of the mass accretion rate with time.

We have confirmed the ERSM conclusion that it is  essentially
impossible to distinguish a highly peaked density distribution (such as
an SIS) from one with a flattened center (such as BE spheres) based on
the millimeter maps.  This has of course no consequences for methods of
determining the dust column density distribution based upon absorption
or extinction measurements (e.g. Alves et al. 2001, Bacmann et al.
2000). These other approaches however have their own limitations.
For example, using NIR colors one is in practise limited to
background giants seen in both H and K which puts an upper limit on
the possible extinctions which are probed. The MIR technique on
the other hand has problems both because one needs to
distinguish between foreground
and background radiation fields and because one is forced to make
assumptions about the uniformity of the background field.
Both these effects make it difficult to provide incontrovertible
evidence of extinctions of order or larger than 100 magnitudes
if present. We believe therefore that
one should still be very cautious about ruling out peaked
density distributions (such as the SIS)
on the basis of the available data.

Using for the first time a reliable dust temperature distribution,
we  have obtained reasonable agreement  between our models based upon
the CB density distribution at time $t_{3}$  corresponding to
a central density of $4\times 10^6$ \percc \ and the available
millimeter and sub-millimeter data for L1544.  This is a strong argument in
favor of the validity of the model of CB as far as L1544
is concerned.  We favor however higher inclinations of L1544 to
the line of sight than given by CB and can fit both the SED and the
aspect ratio for an inclination of 37$^{\circ}$. There is
on the other hand a considerable margin of error in this estimate and
we do not entirely exclude the value of $16^{\circ}$ preferred by
CB.

We conclude in any case that it would be
useful to fit other cores with similar models.  There may be
some possibility in this way of relating the observed structure to
``age'' though it remains to be seen how effective this is in practise.
The high central column density is the best indicator of an age close
to the pivotal situation where dynamical collapse can commence.
Unfortunately, the lower temperature of the high density nucleus
caused by the increased extinction makes this high central column
difficult to discern.

It should be realised also that other models are likely
to be capable of fitting observational constraints. Recent
relevant theoretical  studies include those of Safier, McKee, and
Stahler (1997), Galli et al. (2001) , and Indebetouw and Zweibel(2000).
We note that Galli et al.
(2001) obtain an excellent fit to the observed intensity distribution at 1.3~mm
of L1544 with their model of a slightly inclined ``singular isothermal disk" or
SID.  This is a non--axisymmetric structure with field lines running
approximately in the line of sight.  It would be useful also to compute
the temperature structure of such a model in analogous fashion to that
used in this paper for the CB models.

We have not considered in detail in this study the effect of external
radiation fields other than the ``standard'' fields of MMP
and Black~(1994). However, it is likely that the actual
fields incident upon dense cores vary from core to core and
still more between molecular cloud and molecular cloud. Such
differences plausibly occur due to the presence in the vicinity of
nearby OB stars.  The Bacmann et al.~(2000) study (see their table 1 and
subtract the zodiacal contribution) suggests that there may be as much
as an order of magnitude difference in the ISO LW2 (5-8.5 \mic)
background towards different cores. It in fact also suggests that we
underestimate by a factor of order 3
the field (at 7~\mic) incident on L1544 using the Black~(1994) average
field.  We have examined IRAS maps of L1544 and averaged
intensities
in an annulus surrounding the core with inner radius 100\arcsec \ and outer
radius 300\arcsec \ centred on the mm dust emission peak.
We conclude that the field adjacent to the core is a factor 2.5-4 greater
than the Black~(1994) field at 12 and 25~\mic \ suggesting that we
underestimate the MIR contribution to the incident field (section 3).
This may cause our temperature estimate to be low (by a factor of roughly
1.2) at extinctions of 20-30 magnitudes (see Fig. ~\ref{analytic_temp}).

We may also be under--estimating the FIR field. L1544 is quite clearly
present on IRAS 100\mic \ maps though not present at shorther
wavelengths. This suggests that the incident UV--optical
radiation field (presumably also
responsible for exciting the MIR)  is converted into far IR radiation
in a ``PDR'' layer surrounding the cold core. This presumably should be
added to the average incident field which we have used in our
computations.  We thus may also be underestimating  the FIR heating.

We expect temperatures in these cores to vary as $\chi ^{1/5.6}$
where $\chi $ \ is the scaling factor for the external radiation field
(see section 3) and thus we expect differences as large as 1.5 between
different objects. We would expect for example the nucleus of a core
like OphD on this basis to be roughly 1.5 times hotter than L1544 using
the mid IR intensities of Bacmann et al. (2000).  This will have some
effect on parameters such as the accretion rate but probably not an
important one since the region where gas and dust temperatures are
coupled is very limited. On the other hand, phenomena such as depletion
of molecular species on grain surfaces are extremely sensitive to grain
temperature and these are likely to have different characteristics
depending on the precise grain temperature. This may explain why some
of the chemical characteristics of cores in Ophiuchus and Taurus differ
(long chain carbon species for example are much less prevalent in
Ophiuchus) and may have consequences for the ionization degree
at high densities.

One of the most useful products of this study is the analytic
estimate which we have made of the temperature at the center of a
spherically symmetric core. We have shown that the formula which we
have derived gives in many cases a reasonable fit  to the radial
dependence of dust temperature and we believe it can be used as a useful
first approximation to the gas temperature at densities above $10^5$
\percc . Future studies can then perhaps consider the hydrostatic
equilibrium of ``Bonnor--Ebert spheres'' for a more reasonable temperature
distribution.

A brief summary of the results presented here as well as a more general
discussion of the physical properties of pre--protostellar cores is
given by Walmsley et al. (2001).

\acknowledgement

Many thanks are due to Glenn Ciolek who made available a digital
version of his model predictions for L1544 and to John Black for
sending us a digital version of his estimate for the interstellar
radiation field.  CMW would like to thank the Max Planck Institut
f\"{u}r Radioastronomie for its hospitality  during various phases of
this study.  He also wishes to acknowledge travel support from ASI
Grant ASI-ARS-98-116  as well as from the MURST program
``Dust and Molecules in Astrophysical Environments''. DG acknowledges
financial support from the EC-RTN program ``The Formation 
and Evolution of Young Stellar Clusters'' (RTN1-1999-00436). We would like
to thank Fr\'ed\'erique Motte and Neal Evans for
supplying digital versions of their maps of L1544. Our thanks are also
due to Neal Evans and Antonella Natta for comments on the text.

\appendix

\section{The homogeneous spherical cloud}

Consider a spherical, homogeneous cloud of radius $R$, exposed to
an interstellar radiation field of mean intensity $J^{\rm is}_\nu$.
Let $P$ be an internal point at a distance $r$ from the center
$O$, and define a coordinate system centered in the center of
the sphere with the $z$ axis along $OP$. Let $\theta$ be the angle
between the $z$ axis and a generic direction in space originating from
$P$. Considering only absorption by dust, the mean intensity in $P$ is \be
J_\nu(r)=\frac{J^{\rm is}_\nu}{4\pi}\int_0^\pi e^{-\tau_\nu(r,\theta)}
2\pi\sin\theta\; d\theta, \label{defj} \ee where $\tau_\nu(r,\theta)$
is the optical depth of the cloud at frequency $\nu$ along a
path from the point $P$ to the edge of the cloud in the direction
$\theta$. Since the cloud is homogeneous, $\tau_\nu(r,\theta)$ is
proportional to the distance from $P$ to the surface of the cloud, \be
\tau_\nu(x,\theta)=\tau_{\nu c}(\sqrt{1-x^2\sin^2\theta}-x\cos\theta),
\ee where $\tau_{\nu c}$ is the optical depth at frequency $\nu$
to the center of cloud, and we have set $x\equiv r/R$.  With the
substitution $\eta\equiv \sqrt{1-x^2\sin^2\theta}-x\cos\theta$,
eq.~(\ref{defj}) becomes \be J_\nu(x)=\frac{J^{\rm is}_\nu}{4x}
\left[(1-x^2)\int_{1-x}^{1+x}\frac{e^{-\tau_{\nu c}\eta}}{\eta^2}d\eta
+\int_{1-x}^{1+x}e^{-\tau_{\nu c}\eta}d\eta\right].  \ee Defining the
auxiliary variables $x_\pm\equiv \tau_{\nu c}(1\pm x)$, the integral can
be expressed in terms of exponential-integral functions,
\begin{eqnarray}
\lefteqn{J_\nu(x)=\frac{J^{\rm is}_\nu}{2(x_+-x_-)} \times} \nonumber \\
& & \left[x_+E_2(x_-)-x_-E_2(x_+)+x_-E_0(x_-)-x_+E_0(x_+)\right],
\label{jexact}
\end{eqnarray}
where $E_0(x)=e^{-x}/x$.
The values of $J_\nu(x)$ at the center and at the edge
of the cloud are, respectively
\be
J_\nu(0)=J^{\rm is}_\nu e^{-\tau_{\nu c}},
\label{jcen}
\ee
and
\be
J_\nu(1)=\frac{J^{\rm is}_\nu}{4\tau_{\nu c}}
(1+2\tau_{\nu c}-e^{-2\tau_{\nu c}}).
\label{jedge}
\ee
The value of $J_\nu$ at the edge of the cloud varies between
$\frac{1}{2}J^{\rm is}_\nu$ for $\tau_{\nu c}\rightarrow \infty$ and
$J^{\rm is}_\nu$ for $\tau_{\nu c}\rightarrow 0$.

\section{Parameters for the interstellar field and the opacity}

We have (for the purpose of deriving the analytic formulae given in
section 3.2) made approximate fits to the interstellar radiation field
given by  Black~(1994) as well as to the grain opacities of Ossenkopf
\& Henning~(1994).  We approximate the radiation field by a sum of
modified Planck functions given by eq.~(\ref{rfield}) and the opacity
by piecewise power laws as follows.

For each black body (or modified) component of the interstellar
radiation field, we give in Table ~\ref{isf_param} the values of the
parameters $\lambda_p$, $p$, $W_i$, $T_i$ discussed in section
3.2.

The opacity parameters $\alpha$ and $Q_\nu$ for the different frequency
ranges  are also given in Table ~\ref{op_param}.  These are based on
the fact that one gets a rough fit (20 percent) to the Ossenkopf \&
Henning~(1994) opacity results with a piecewise power law fit having an
index 1.4 below 10~\mic, 1.6 between 10~\mic\ and 400~\mic, and 2
longward of 400~\mic\ (opacity proportional to $\nu^\alpha$).  The
parameters for dust emission ($Q_e$ and $\alpha_e$) used in the
analytic formulation are appropriate for dust grains at roughly 10~K
and we therefore use the values given for the wavelength range
10--400~\mic\ in Table ~\ref{op_param}.

\begin{table}
\caption{\sc Parameters for the interstellar field}
\vspace{1em}
\begin{tabular}{lllll}
\hline
range        & $\lambda_p$  & $p$  &  $W_i$  & $T_i$ \\
             &                    &      &         & ($K$) \\
\hline
V--NIR       & 0.4~\mic\   &  0   &  $1\times 10^{-14}$  & 7500 \\
             & 0.75~\mic\  & ''   &  $1\times 10^{-13}$  & 4000 \\
             & 1~\mic\     & ''   &  $4\times 10^{-13}$  & 3000 \\
\hline
MIR          & 100~\mic\   &$-1.8$&  $5\times 10^{-7}$   &      \\
\hline
FIR          & 140~\mic\   & 1.65 &  $2\times 10^{-4}$   & 23.3 \\
\hline
CBR          & 1.06~mm     & 0    &  1                   & 2.728 \\
\end{tabular}
\label{isf_param}
\end{table}

\begin{table}
\caption{\sc Parameters for the opacity}
\vspace{1em}
\begin{tabular}{llll}
\hline
wavelength range   & $\lambda_{\rm a}$  & $Q_\nu$ at $\lambda_{\rm a}$ & $\alpha$  \\
                   &                    & (cm$^2$~H$_2^{-1}$)          &           \\
\hline
0.1--10~\mic\   & 1~\mic\    & $3.9\times 10^{-22}$ & 1.4 \\
10--400~\mic    & 140~\mic\  & $1.5\times 10^{-24}$ & 1.6 \\
400~\mic--10~mm & 1.06~mm    & $3.3\times 10^{-26}$ & 2.0 \\
\end{tabular}
\label{op_param}
\end{table}

\bibliography{}
\bibliographystyle{astron}

\end{document}